\documentclass[prb,reprint,superscriptaddress]{revtex4-1} 

\usepackage{amsmath}  
\usepackage{amsfonts} 
\usepackage{graphicx} 
\usepackage{booktabs} 
\usepackage{colortbl} 
\usepackage{xcolor} 
\usepackage{xfrac}
\usepackage{enumitem}

\begin{document}


\title{Student understanding of electric and magnetic fields in materials}

\author{Savannah L. Mitchem}
\email{These authors contributed equally to the work of this manuscript and are listed alphabetically.}
\address{Department of Physics, Florida State University, Tallahassee, Florida, 32306, USA}

\author{Dina Zohrabi Alaee}
\email{These authors contributed equally to the work of this manuscript and are listed alphabetically.}
\address{Department of Physics, Kansas State University, Manhattan, Kansas 66506, USA}

\author{Eleanor C. Sayre}
\email{esayre@ksu.edu}
\address{Department of Physics, Kansas State University, Manhattan, Kansas 66506, USA}

\date{\today}

\begin{abstract}
We discuss the clusters of resources that emerge when upper-division students write about electromagnetic fields in linear materials. The data analyzed for this paper comes from students' written tests in an upper-division electricity and magnetism course. We examine how these clusters change with time and context. The evidence shows that students benefit from activating resources related to the internal structure of the atom when thinking about electric fields and their effect on materials. We argue that facilitating activation of certain resources by the instructor in the classroom can affect the plasticity of those resources in the student, making them more solid and easily activated.  We find that the wording of the questions posed to students affects which resources are activated, and that students often fill in resources to link known phenomena to phenomena described by the question when lacking detailed mental models.

\end{abstract}
\maketitle 

\section{Introduction}

Learning to think scientifically is extremely important for all students.  All science, technology, engineering, and math (STEM) majors must take physics; thus, research into how students learn physics is important for all STEM students. A significant body of research in physics education has been devoted to improving the teaching and evaluation\cite{Maloney2001} of electricity and magnetism (E\&M) at the introductory level\cite{Singh2005Gauss, McDermott1999Resource, Chabay2006Restructuring} but less attention has been geared towards upper division E\&M. \cite{Manogue2006Ampere, Bilak2007, Bing2007Blending, Pepper2010Gauss, Traxler2007Gauss, Wallace2010} Learning physics at the upper division level is a complex endeavor, and each student is unique in the way they process information to understand upper level physical concepts. Prior research into student learning of E\&M at the upper division level has focused on student understanding of specific topics like Maxwell's equations. \cite{Pepper2010Gauss, Traxler2007Gauss, Wallace2010, Baily2013, Baily2015} Other research uses E\&M as a context in which to study mathematical tools which support physical reasoning, \cite{Wilcox2014, Pepper1012Observations} student identity development,\cite{Sayre2015BESM} or epistemological framing.\cite{Nguyen2016Dynamics} Of the studies that focus on student understanding of specific physics topics, none focus on how students understand the behavior of linear materials in electric and magnetic fields. This is an important topic to research in science and technology. For instance, dielectrics have numerous practical applications in the home and industry, including: nanotechnology, electronics, photonics, chemical and mechanical systems, and the emerging fields of biology and biochemistry. So, we need to help students develop mechanistic reasoning while covering these topics that are fundamental to many STEM fields.

The development of mechanistic reasoning plays an important role in student understanding, and evaluating understanding in STEM areas often involves unpacking the structure of student reasoning. We use Resource Theory\cite{Hammer2000} to describe the ways students link pieces of information, or resources,  to form more complex ideas, improve understanding, and solve problems. As researchers gain more insight into the diversity of student thought processes, patterns in how students use and combine resources help us better understand student learning. In this paper, we discuss student responses to conceptual questions about linear materials in electric and magnetic fields.  We categorize the responses by the students' resource usage, and we look for changes across contexts and over time.

\section{Theory}

Previous research on Resource Theory and mental models in Electricity and Magnetism guides the research described in this paper. A resource, broadly speaking, is a discrete piece of an idea that a student activates when considering or solving a problem.\cite{Hammer2000}  It is important to note that resources themselves are not inherently right or wrong (e.g. \textit{closer means stronger}),\cite{diSessa1993} and that student difficulties arise from misapplying resources (e.g. it's hotter in the summer because the Earth is closer to the sun). Students build resources throughout their lives and education,\cite{Sayre2008} and they use these resources by activating and linking them to form mental models of complex physical and mathematical phenomena.  Unlike phenomenological primitives, or p-prims for short, which are the simplest, most atomic pieces of knowledge,\cite{diSessa1993} resources can have internal structure that is accessible to the user.\cite{Wittmann2006Conceptual,Sayre2008} Because resources can be more complex and structured, students can develop resources for sophisticated topics such as diode construction,\cite{Sayre2003} separating differential equations,\cite{Wittmann2015,Black2009} or quantum mechanics.\cite{Gire2008Resources}

Several kinds of resources have been identified.   Conceptual resources are pieces of knowledge, such as \textit{coordinate systems}\cite{Sayre2008} or \textit{activating agent}.\cite{diSessa1993}  Procedural resources\cite{Black2009} are actions such as bringing constants out of a derivative\cite{Modir2016Pulling} or summing forces.  Epistemological resources\cite{Hammer2003Tapping} are elements of beliefs about the nature of knowledge, such as whether results can be figured out or if they need to be looked up. To solve problems, students coordinate all three kinds of resources, activating connections between conceptual resources and procedural ones to build arguments, as mediated by epistemological resources about the problem solving they are engaged in. We are specifically interested in conceptual resources, and how instructors can recognize the resources students bring to the course and how they can encourage the use of those resources as applied to electricity and magnetism.\cite{Chabay2006Restructuring} We explore in this paper which conceptual resources are activated in clusters when students explain polarization and magnetization. 

Another characteristic of resources is their plasticity versus their solidity.\cite{Sayre2008} Solid resources are durable and well established, and their internal structure need not be accessed for  use. In other words, they require less justification and can be more easily and readily used and linked with other resources. Plastic resources are unstable and require more elaboration and justification to be used. They are usually new resources for the student, and their connections to other resources have yet to be solidified. More broadly, ``stable networks'' are networks of resources that are ``cognitively nearby,''\cite{Sayre2003} and tend to activate together.\cite{Sayre2008} In this paper, we examine how the plasticity of conceptual resources and the networks and clusters of resources students use change with increased conceptual instruction.

Solid networks of resources form students' mental models. Mental models ``enable individuals to make predictions and inferences, to understand phenomena and events, to make decisions, and to control their execution. They are incomplete, despite being structural analogues of the processes taking place in the world''.\cite{Borges1999Electricity} After interviewing students across a wide range of ages and backgrounds, Borges and Gilbert \cite{Borges1998Magnetism} gained insight into mental models of electricity, and found that the more successful and complete the model was, the more microscopic (very detailed) it tended to be. They also discuss students' use of causal agents in their reasoning. They discovered that many of the more incomplete mental models the subjects had about electric current moving through a wire were characterized by the interchangeability of words such as ``electricity'', ``current'', and ``energy''. The subjects knew that a causal agent must come between perceived related events, such as a wire being connected and a light bulb lighting up, and they filled in resources about ``something'' flowing, without knowing what that ``thing'' was.

Borges and Gilbert investigated mental models of magnetism as well. Although their discoveries were of mental models of magnets creating magnetic fields, those models may be closely related to the mental models students have of materials subjected to a magnetic field. Again, the most complete mental models (meaning the ones with the most power in making predictions and inferences about a situation)\cite{Borges1998Magnetism} were also the most microscopic models (very detailed). They also found that the participants interviewed with the most prior experience and formal instruction about magnetism were the only ones who thought of magnetism as a result of micro-currents circling around the atoms in a material that line up to produce a field. They concluded that deliberate instruction can affect students' mental models, noting that ``when comparing students' models, the effects of their models are evident, as shown by the vocabulary and the type of constructs they use.''\cite{Borges1998Magnetism}

It is important to note that it is possible to have a complete mental model that is applied incorrectly. For example, some participants in our study showed evidence of having a complete mental model of polarization of light that they applied incorrectly to our question about polarization of materials. It is also possible to have an incomplete mental model that is applied correctly. For example, a student could make correct predictions about the results of a physical situation, but may be only able to do so for very specific cases.  In this paper, we're interested in both students' answers and the reasoning they use to generate them.

\section{Methods}
\subsection{Data collection}
The data analyzed for this paper comes from students' written test responses spanning two years at Kansas State University. The upper-division E\&M course is available every fall semester, and typically covers the first half of Griffiths's text,\cite{Griffiths1999EM} Maxwell's equations in matter and vacuum. The class meets four days per week for fifty minutes at a time.  About one third of the days are tutorial days in which students complete guided practice problems in small groups instead of listening to a lecture. 

There were two cohorts of students who took the class in 2014 and 2015. In 2014, on the final exam (``Final 2014''), 19 students were asked both parts (1) and (2) from Figure ~\ref{Fig1}.  In 2015, 18 students were asked only part (2) of this problem (electric fields) near the beginning of instruction on this topic in the 12th week of the semester (``Mid-Semester 2015'', we treated this as a pre-test), and all parts of the problem on their final exam (``Final 2015''). Between the 2015 pretest and final, students were given targeted instruction on this problem. Thirty-eight students in total completed the class in 2014 and 2015, and 55 assessments (written test responses) were collected, both during instruction and at the final exam. We analyzed all usable answers to find patterns and trends in the students' use of resources, and we also searched for patterns among the 2015 students' responses during the semester and after they completed the class.

\begin{figure}[tbh]
\fbox{%
\begin{minipage}[l]{0.9\linewidth}%
Linear materials encounter strong fields, and then linearity breaks down.  \textit{Hint: You don't have to do any quantum here! Our ordinary assumptions about matter are sufficient.}
\begin{enumerate}
\item In a sufficiently strong external magnetic field, a linear material cannot become more magnetized.  That is, $\mathbf{M} = \chi_m \mathbf{H}$ holds until some threshold value of $\mathbf{H}$, above which $\mathbf{H}$ may increase but $\mathbf{M}$ cannot.  Explain, in words and pictures as appropriate:
\begin{enumerate}
\item why this happens; and,
\item what happens in the material when the threshold value of $\mathbf{H}$ is exceeded.
\end{enumerate}
\item In a sufficiently strong external electric field, a linear material cannot become more polarized.  That is, $\mathbf{P} = \epsilon_0 \chi_e \mathbf{E}$ holds until some threshold value of $\mathbf{E}$, above which $\mathbf{E}$ may increase but $\mathbf{P}$ cannot.  Explain, in words and pictures as appropriate:
\begin{enumerate}
\item why this happens; and,
\item what happens in the material when the threshold value of $\mathbf{E}$ is exceeded.
\end{enumerate}
\end{enumerate}
\end{minipage}%
}
\caption{Final exam question for 2014 and 2015 cohorts.  At mid-semester in 2015, the students were quizzed only on part (2).}
\label{Fig1}
\end{figure}

\subsection{Data analysis}

We reviewed all 55 student exams, seeking patterns in student understanding. Three students did not answer the question and two students wrote about the unrelated phenomenon of polarization of light. We discarded these students' exams from further analysis. In addition, one student did not take Final 2015 and one student was absent for Mid-Semester 2015, leaving 50 usable responses in all, 33 for 2015 and 17 for 2014. 

To find patterns, we considered each response's resources, drawing resource graphs\cite{Wittmann2006Conceptual} to generalize them across responses. We use resource graphs to link resources activated in particular contexts. As themes emerged in the data, we sought a categorization scheme which would both capture the rich variety of student responses and make meaningful categorical differences among them. We iteratively developed an emergent coding scheme linking specific student language to specific resource use.


We classified each response as \textit{sub-atomic} or \textit{super-atomic} (which are explained in the next section). Then we noted which resources each student used by the criteria in our codebook. If the student's resources did not fall into any of the categories we defined, we marked them as ``other.''

When our coding scheme was stable, we invited an additional researcher to separately code student responses.  He categorized 20 randomly selected answers, using our categorization scheme for evaluation during inter-rater reliability testing (IRR). Before the discussion, 82\% of his identified resources and ours overlapped. After discussion, which included clarifying and improving parts of the codebook and discussing each response categorized differently, we repeated this procedure with 20 different responses and found 100\% agreement.

\section{The problem and its solution}

The following is a complete and correct response to the exam question students were asked. When a dielectric material is placed in an electric field, its atoms become polarized. This means that the negative electron cloud around the positive nucleus is slightly displaced from the center. This displacement of charges within the atoms forms electric dipoles with separation linearly proportional to the strength of the electric field, as given by the equation $\mathbf{P} = \epsilon_0 \chi_e \mathbf{E}$ where $\mathbf{P}$ is net polarization, $\mathbf{E}$ is the electric field, $\chi_e$ is a property of the material (susceptibility), and $\epsilon_0$ is constant. When the electric field's strength exceeds a certain threshold value, electrons become separated from their nuclei, converting the dielectric into a conductor. This is called dielectric breakdown.  

When a linear paramagnetic material is placed in a magnetic field, whatever magnetic dipoles that already exist become more aligned with the field, tending to point in the same direction as each other. As opposed to polarization, magnetization of a material does not separate any charges, but merely changes the random direction of spins within the atoms to be more uniform. The magnetization $\mathbf{M}$ of a material is given by $\mathbf{M} = \chi_m \mathbf{H}$ where $\mathbf{H}$ is the magnetic field and $\chi_m$ is a constant dependent on the material. The linear relationship between $\mathbf{M}$ and $\mathbf{H}$ breaks down after $\mathbf{H}$ is strong enough to align all of the magnetic moments of the atoms parallel to it. After that threshold, $\mathbf{H}$ can increase but $\mathbf{M}$ cannot. No breakdown of the material occurs.

\section{Results}
\subsection{E-Field}

As part of our investigation, we looked at how students considered what happens to a dielectric material in an extremely strong electric field. We examined their responses for evidence of resources being activated. When observing the students' answers to the polarization question, we found that the resources students activated generally fell into two groups. The first group, which we call \textit{sub-atomic}, consists of students who used resources related to the internal structure of the atoms within the material. This includes mention of electron clouds along with nuclei, stretching of the negative and positive charges within the atom, drawings of dipoles with positive and negative sides, and internal forces of the atom versus the external force of the electric field. 23 of 50 responses are in this category. 

The second group, \textit{super-atomic}, consists of students who, instead of thinking about the internal structure of the atom, talked only about the material in general. These students often considered the problem in terms of the model of positive and negative charges moving throughout the material and collecting on one side, as opposed to thinking about the atoms themselves becoming dipoles aligned with, and pulled apart by, the field. Included in this category are drawings of material with pluses and minuses lined up within it (that do not form polarized atoms), mention of electrons or charges without reference to an atom, and mention of the word ``dipole'' without description or depiction of its structure. 27 of 50 responses are in this category. 

When comparing the answers of the \textit{sub-atomic} and \textit{super-atomic} groups represented in Table~\ref{table1}, we found that if a student considers the internal structure of the atom during polarization, they are very likely to arrive at the correct answer, dielectric \textit{breakdown}. This is true for all three testing situations we evaluated. Furthermore, on the two tests where students had not been given the question before (Mid-Semester 2015 and Final 2014), the students that took the \textit{super-atomic} route were not likely to get to \textit{breakdown} over other answers. This is especially true for Mid-Semester 2015, when the students had not yet been taught the material. 

\begin{table}[tbh]
\caption{Resource groups activated by students versus their answers while writing about material in fields.}
\begin{ruledtabular}
\begin{tabular}{l c c}
\toprule 
  & \textit{Sub-atomic} & \textit{Super-atomic} \\
\hline 
\textbf{Final 2014} & & \\
Break down	& 5	& 3 \\
Saturation	& 1	& 6 \\
Others	& 0	& 2 \\
\hline 
\textbf{Mid-Semester  2015}\\
Break down	& 7	& 1 \\
Saturation	& 0	& 7 \\
Others	& 0	& 1 \\
\hline 
\textbf{Final 2015}\\
Break down	& 9	& 4 \\
Saturation	& 1	& 2 \\
Others	& 0	& 1 \\

\end{tabular}
\end{ruledtabular}

\label{table1}
\end{table}

\begin{figure*}[tbh]
	\centering
	\includegraphics[width=0.8\linewidth]{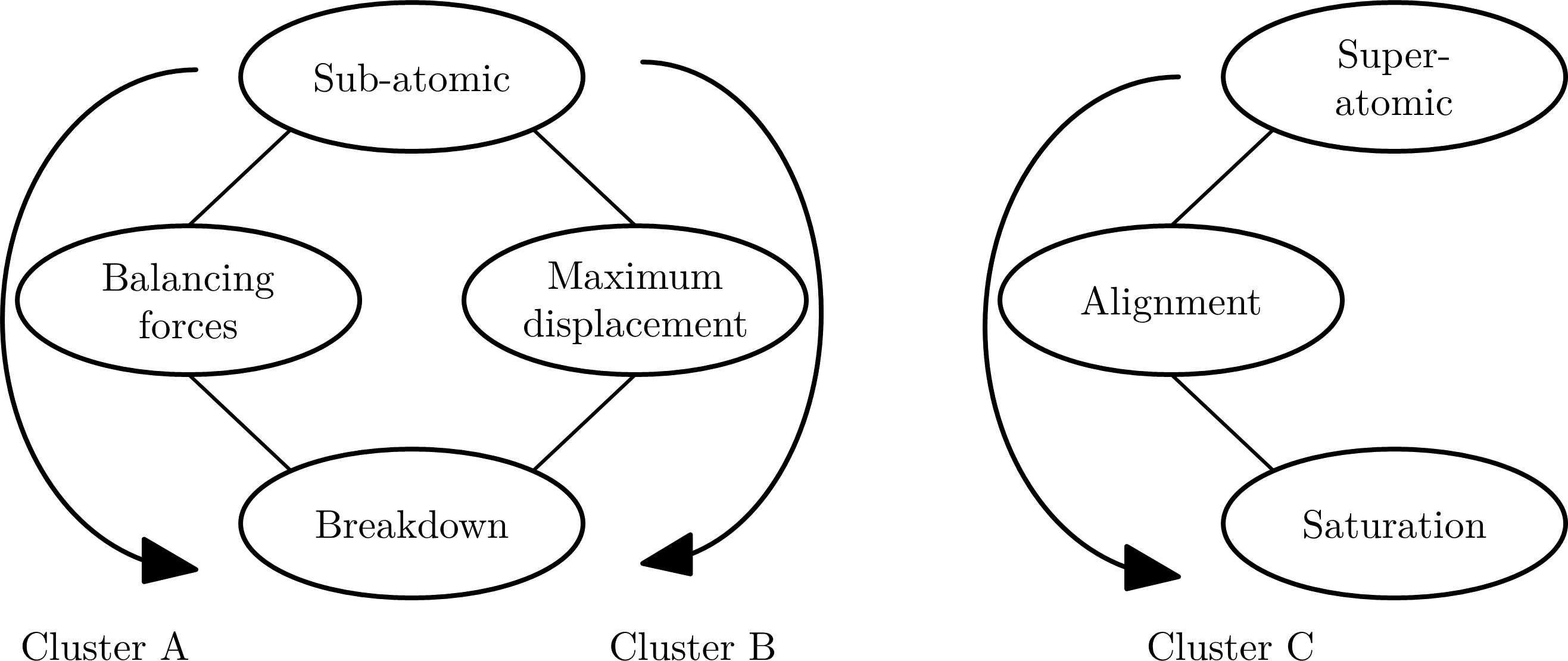}
	\caption{Resource graph for materials in electric fields. The diagram on the left represents Clusters A and B, and the diagram on the right represents Cluster C.}
	\label{Fig2}
\end{figure*}

There were other variations of resources used within the large groups of \textit{sub-atomic} and \textit{super-atomic} resources. Students who used \textit{sub-atomic} resources and reached \textit{breakdown} tended to justify their answers one of two ways. In the first approach, students described the internal atomic forces holding the electrons and nuclei together, and how an electric field above a certain threshold would provide a large enough force to overcome the internal force and break electrons away from their atoms. We called this resource \textit{balancing forces}, and Cluster A is the name we gave the cluster of resources including \textit{sub-atomic} resources, \textit{balancing forces }, and \textit{breakdown}. In the second method, students described how maximum polarization pulls the electron cloud so far away from the nucleus that electrons break from their nuclei. We called this resource \textit{maximum displacement}, and Cluster B is the name we gave the cluster of resources including \textit{sub-atomic} resources, \textit{maximum displacement}, and \textit{breakdown}. They are essentially the same phenomenon, but \textit{balancing forces} is a more rigorous description. Figure ~\ref{Fig2} depicts the common clusters of resources used by students. 

An example of a student response that includes \textit{sub-atomic} resources and the \textit{balancing forces} resource (Cluster A) is that of ``Kate''\footnote{All student names are pseudonyms.} in Mid-Semester 2015. ``Polarizing atoms means slightly offsetting the negative electron cloud from the positive nucleus.'' It is clear that Kate uses \textit{sub-atomic} resources to frame the question. She goes on to say, ``If the field doing this is strong enough, it will separate the two entirely, since it will overcome the Coulombic force attracting the electrons to the nucleus. The electrons will gather on one side of the material.'' The use of the words ``overcome'' and ``force attracting the electrons to the nucleus'' shows that \textit{balancing forces} activated along with \textit{breakdown}. 

In her answer for Final 2015, Kate used Cluster B instead. She wrote, ``The external electric field pulls on the electron clouds in the individual atoms/molecules, creating a polarization.'' She mentioned electron clouds in relation to their atoms, so we classified her as using \textit{sub-atomic} resources, seeing as she conceptualized the situation on a sub-atomic level.  Kate continued, writing ``As the field increases, the electron clouds get pulled farther and farther away from the nucleus, until they can go no farther without leaving the nucleus entirely.'' We classified this resource as \textit{maximum displacement} because she mentioned the distance between the electron clouds and the nuclei reaching a breaking point. For part two of the question, she wrote, ``The electrons are ripped away from the nuclei, causing a brief flow of charge (dielectric breakdown)''.

Students who did not think about polarization at the sub-atomic level tended to justify their answers using a description of the dipoles or charges aligning with the field until none of the dipoles or charges could be more aligned. This resource, \textit{alignment}, was usually paired with the final answer of \textit{saturation}. These students wrote that nothing significant happened to the material besides \textit{alignment}, failing to mention \textit{breakdown}. We called this cluster of resources Cluster C which included \textit{super-atomic} resources, \textit{alignment}, and \textit{saturation}. Several students' responses (fifteen responses in all) that used this cluster made no mention of separation of charges during polarization and only mentioned \textit{alignment} of the dipoles, treating polarization and magnetization as the same phenomenon in their answers. Many students used a model of charges aligning in some way across the material while separating as well.

An example of Cluster C is Alex's answer in Final 2014. He wrote, ``Polarization occurs when an E-field pushes all $+$ charges one way and the $-$ charges move the opposite direction. This causes an E-field from the difference between separated $+$ and $-$ charges.'' There is no mention of atoms here, and the term ``charges,'' especially when he discusses positive charges, indicates his use of the model of charges moving throughout an insulator and causing net effects rather than electrons moving within an atom. He went on to write, ``Eventually all the $+$ will be as far to one side as possible and all of $-$ will be as far as possible in the opposite direction. The charges are as far away as possible with all the charges segregated. At this point, no more charges can move so E will increase, but P can't.'' Here, he activates the resources \textit{alignment} and \textit{saturation}. In summary, he uses \textit{super-atomic} resources, \textit{alignment}, and \textit{saturation}. 

The rest of the students' answers indicated the activation of other resource clusters which we grouped into the miscellaneous Cluster D. This includes any answers that were neither \textit{breakdown} nor \textit{saturation}. It also includes students who used \textit{sub-atomic} resources but reached an answer of \textit{saturation}, or students who used \textit{super-atomic} resources but reached an answer of \textit{breakdown}. And finally, it includes the few students who used none of these or misunderstood what the question asked of them. 

The grouping of these resources into clusters aided us in finding patterns in student responses over varied context and amount of instruction. As stated before, for all three testing situations, using \textit{sub-atomic} resources almost always coincided with activating the resource of \textit{breakdown}. Conversely, activation of \textit{super-atomic} resources typically coincided with activation of the \textit{saturation} resource instead of \textit{breakdown}. Interestingly though, on Final 2015, there was not as much of a difference between the number of \textit{sub-atomic} and \textit{super-atomic} students who arrived at \textit{breakdown}. In fact, 13 out of 17 students reached \textit{breakdown} on Final 2015, whereas only 8 out of 16 reached \textit{breakdown} on Mid-Semester 2015. Below are details about why this may have occurred. 

Table~\ref{table2} summarizes the number of students who used each Cluster on Mid-Semester 2015 and Final 2015. Four students switched from using Cluster A, B, or C in the pretest to using Cluster D (and still getting \textit{breakdown}) on the final. Take Zachary's response for Final 2015 for which he wrote, ``The material is bounded. Above and below the faces of the material are regions of different $E_r$ and $\chi_e$. These regions do not have the same property as the linear material, and so the max P occurs at E threshold, when all the charges are on opposing faces.'' Along with his written answer, Zachary also drew a diagram of a rectangular prism in an electric field. He went on to say, ``To remain in equilibrium, the dipole $E_i=E_{ext}$. This is achieved by highly polarizing the material, so all the positive charges are on top face, and $-$ charges on bottom face. This occurs at E threshold. If $E_{ext} > E_{threshold}$, the charges can't balance out, and the charges begin to strip away from the material and end the linear property.'' 
 
 Zachary was able to reach \textit{breakdown} without activating \textit{sub-atomic} resources, unlike in his answer for the question during the pretest for which he used Cluster A. He discusses only \textit{super-atomic} qualities and phenomena in his response after instruction about materials in electric fields and the internal process of polarization. This could be due to the fact that the instructor facilitated activation of those \textit{sub-atomic} resources in a way that allowed them to go from being plastic to solid in Zachary's mind. The instructor reviewed the exact question after the pretest, and used \textit{sub-atomic} descriptions in a Mid-Semester lecture including the fact that each atom is made up of a little bit of negative and positive, and the positive center is surrounded by the negative outside. She also describes \textit{breakdown} as when ``electrons get ripped away from atoms,'' and contrasts separation and alignment. This lecturing could have increased Zachary's ability to use \textit{breakdown} as a resource on the final exam without accessing its internal structure or having to justify it further by using \textit{sub-atomic} resources.
\begin{table}[tbh]
	\caption{Comparing the cluster of resources during Mid-Semester and Final 2015 }
	\begin{ruledtabular}
		\begin{tabular}{l c c}
			\toprule \\ 
			& Mid-Semester test & Final test \\
			\hline  
			Cluster A	& 4	& 7 \\
			\hline 
			Cluster B	& 3	& 2 \\
			\hline 
			Cluster C	& 7	& 2 \\
			\hline 
			Cluster D	& 2	& 6 \\
		\end{tabular}
	\end{ruledtabular}
	\label{table2}
\end{table}

Four other students used Cluster C or D, getting \textit{saturation}, on the pretest, and then switched to Cluster A or B on the final, indicating that they understood the lecture and used the resources given to them by instruction to correctly answer the final. (Note that when we use the word "correctly" here, we are referring only to the final answer and are not commenting on the correctness of the reasoning the student used.) The remaining students either used the same cluster for both exams, or used a different cluster resulting in the same answer for both. 

\subsection{B-field}

The second part of our research was to analyze data from the portion of the question on the two final exams about materials in extremely high magnetic fields. This part of the question was not included in Mid-Semester 2015. Unlike the polarization question, none of the students who answered the magnetization question had seen the problem before their final exam. However, they had all been taught concepts about magnetic fields and magnetization through practice problems and lectures prior to the final. The purpose of this question was to get students to compare and contrast what happens to linear materials when they are highly polarized versus highly magnetized. 

The vast majority of student responses included only \textit{super-atomic} resources to describe and explain magnetization and its effect on the material. These resources consisted mainly of alignment of atoms in some sense. Only two students did not reach the correct answer of \textit{saturation}, and interestingly, those students used \textit{sub-atomic} resources in their responses. Those students mentioned the internal structure of the atom during magnetization, and they concluded with \textit{breakdown} of the material. 

\textit{Breakdown} and \textit{saturation} are the correct answers for E-field and B-field, respectively. So by switching the context, the same resource can be applied correctly or incorrectly. That is, resources by themselves are neither right nor wrong.

\subsection{Comparison between class activities and student responses}

We recorded the class meetings in 2015 about magnetization as part of a broader project on student reasoning during class.  Upon reviewing the lectures, we discovered how different the magnetization lectures were from the polarization lectures discussed in the last section. The instructor used almost no conceptual or \textit{sub-atomic} description of magnetization, and instead focused on bolstering the students' mathematical understanding of the way material behaves in high magnetic fields. She mentioned ways different materials hold magnetization and for how long (such as the ferromagnet's ability to become a permanent magnet), and several students included that information in their responses. However, the only conceptual explanation on a \textit{sub-atomic} level was that atoms can 'align with or against the field based on the thing that is doing the spinning to make the current in the dipole.' She says that this ``thing'' is ``all kinds of quantum-y things'' and has to do with how electrons are paired in atoms. 

This is a contrast from the lecture on polarization which focused on a \textit{sub-atomic} description of \textit{breakdown}. The student responses reflect this difference. Although only two students said that something happens to the material besides maximum alignment (and therefore \textit{saturation}), the ``things'' that students said were aligning varied widely. Over twenty different words were used for this ``thing,'' and most of the pictures drawn were arrows lining up with each other, indicating a \textit{super-atomic} model of magnetization. The words used included correct terminology such as ``magnetic moment'', ``dipole'', and ``electron spin'', and incorrect or incomplete terminology such as ``electrons'', ``atoms'', and ``little pieces.'' All answers including \textit{alignment} and \textit{saturation} were given credit. This lack of a complete group of resources to make up a mental model of magnetization could be due to the instruction the students were given. The students lacked a link between the idea that material is made of atoms and that something about those atoms aligns during magnetization. A similar result was found in Borges and Gilbert's research in which they describe student use of a `causal agent' (in this case, the "thing" that is aligning) to fill in logical gaps left by the question. \cite{Borges1998Magnetism}  Furthermore, previous work on how students understand electric and magnetic fields together suggest that students often confuse these fields or their effects with each other.\cite{Heckler2010, Sayre2009, Scaife2011}For example, Scaife and Heckler found that interference between electric and magnetic concepts can make students confused. They also stated that students' responses depend on whether electric or magnetic force questions are posed first, and this effect depends on whether electric or magnetic force was most recently taught.\cite{Scaife2011}

The wording of the question may have also played a role in the high number of \textit{saturation} responses. Although both questions about the electric and magnetic fields are worded the same, the data should not be analyzed independently from the questions asked. This is because the nature of the response expected of the student for each question was different, and the question itself can activate certain resources. The phrase ``cannot become more magnetized'' in the question is essentially the definition of saturation, and this may have activated resources related to \textit{saturation}. The students could have then activated resources to link their knowledge of magnetization with the idea of \textit{saturation}. In conclusion, wording of this part of the problem may aid students to find the correct answer, \textit{saturation}, and help them to obtain full credit.

Similarly, the polarization question uses the phrase ``cannot become more polarized.'' This phrase may have activated \textit{saturation} resources in some of the students as well. However, for the polarization part, the instructor expected students to take their mental models of materials in electric fields and extrapolate them to explain a special case, breakdown. They are expected to reason further than the default of \textit{saturation} set by the question. In other words, although the wording of the question was identical for both magnetization and polarization, it favored \textit{saturation} as opposed to \textit{breakdown}. Borges and Gilbert's research also supports the claim that student responses and explanations should not be analyzed separately from the question they were asked.\cite{Borges1998Magnetism}

\section{Summary \& conclusion}

We discussed the clusters of resources that were activated together when upper division E\&M students expounded on the behavior of materials in fields. Some examples show that those clusters can change with time and context, specifically with increased instruction of the concepts in question. We have offered insight to which mental models are most helpful for fostering complete conceptual understanding of polarization and magnetization. 

The data indicate that use of resources related to the internal structure of the atom during polarization increased likelihood of activating \textit{breakdown}. Students who did not activate these \textit{sub-atomic} resources tended to activate \textit{saturation} instead. This predictive relationship was not as strong on Final 2015, where we argue that because the students had seen the question and received instruction, the resource clusters they used changed from Mid-Semester 2015. The \textit{breakdown} resource became more solid and required less justification for use, making the students' answers less \textit{sub-atomic} but not less correct.

We also assert that wording of a question has a role in activating certain resources in the student and can induce mechanistic reasoning.\cite{Russ2008} By reading the question, students knew that some aspect of a material becomes saturated when it is in an extreme magnetic field, and they filled in objects to link their knowledge that material is made up of ``things'', and that these things align in the presence of the field.

This research yields multiple implications for instruction of E\&M. After comparing the answers in Part A and B of the final question and watching videos of the lectures corresponding to them, we hope to encourage instructors to put more emphasis on \textit{sub-atomic} conceptual models when teaching polarization. These implications for instruction can also apply to lower and higher level courses than upper division E\&M, and conceptual understanding of physical phenomena on a \textit{sub-atomic} scale could be a helpful supplement to math based curricula as well.

More broadly, when considering how resources are activated and whether these resources are more solid or more plastic, the general goal of instruction is to help resources become more solid. The reasoning attached to solid resources is more complete and more well-connected than that of plastic resources. However, there are two caveats. First, as resources become more solid, students don'€™t need to look inside anymore. This is not necessarily a negative effect, but it could be negative in a circumstance where accessing internal structure would benefit the student. Second, occasionally resources are misapplied and become solid in their incorrect usage and this is where deep-seeded misconceptions come from. \cite{Sayre2008} 

In conclusion, the goals of this study were to identify the clusters of resources that successful students activated while answering an upper-level conceptual problem in E\&M, and to highlight the role of the instructor to facilitate activation of those resources.These results should improve our understanding of how students reason about fields in materials and could yield insight into instructional strategies to improve the learning and teaching of physics at upper division level. 

We know that active learning environments paired with conceptual curricula improve student understanding of physics, but improved instructional methods are still needed to support students' reasoning in these areas, especially for upper division courses. Finally, our work highlights the need for future research on students' use of resources to solve problems in these upper division physics courses. We believe that thinking about polarization inside the atom seems to increase understanding and can give students better intuition about special cases such as dielectric breakdown. Materials with dielectric properties are crucial to modern science, technology, and manufacturing, so improving student understanding of them is important. 

\begin{acknowledgments}

We would like to thank the Kansas State University Physics Education Research group (KSUPER) for assistance on this paper and for helpful feedback. In particular, we would like to thank Dean Zollman for reviewing the manuscript and providing feedback. We would also like to thank Hai Nguyen for his assistance with the IRR. Portions of this research were supported by the KSU Department of Physics, NSF grants DUE-1430967 and PHYS-1461251, and the Air Force Office of Scientific Research.  

\end{acknowledgments}


%

\end{document}